\documentclass[11pt, letterpaper]{article}

\usepackage{ifthen}

\ifthenelse{\isundefined{\MyMacro}}{

\newcommand{\MyMacro}[4]{
\newboolean{#1}
\setboolean{#1}{#2}
\newcommand{#3}{\ifthenelse{\boolean{#1}}}
\newcommand{#4}{\ifthenelse{\not \boolean{#1}}}
}

}

\newcommand{\ifndef}[2]{\ifthenelse{\isundefined{#1}}{#2}{}}

\newcommand{\mydef}[2]{\def#1{#2}}

\newcommand{\nospell}[1]{#1}   %

\usepackage{amssymb}
\usepackage{amsmath}   %

\usepackage{amsthm}    %
\usepackage{latexsym}
\usepackage{amsfonts}
\usepackage{units}     %
\usepackage{psfrag}    %
\usepackage{url}    %

\usepackage[dvips]{graphicx}
\usepackage[usenames,dvipsnames]{color}

\newcommand{\MyComment}[1]{\ClassWarning{My Macros}{#1}}

\ifndef{\theorem}{\newtheorem{theorem}{Theorem}[section]}
\ifndef{\lemma}{\newtheorem{lemma}[theorem]{Lemma}}
\ifndef{\corollary}{\newtheorem{corollary}[theorem]{Corollary}}
\ifndef{\remark}{\theoremstyle{remark} \newtheorem{remark}{Remark}}
\ifndef{\proposition}{\newtheorem{proposition}{Proposition}}
\newtheoremstyle{mydefinition}   %
{\topsep}{\topsep}   %
{\slshape}   %
{}   %
{\bfseries}   %
{.}   %
{ }   %
{}   %
{\theoremstyle{mydefinition}}
\newtheoremstyle{myexample}   %
{\topsep}{\topsep}   %
{\itshape}   %
{}   %
{\slshape}   %
{:}   %
{ }   %
{\ul{\thmname{#1}}}   %
\ifndef{\example}{\theoremstyle{myexample} \newtheorem{example}{Example}}
\ifndef{\claim}{\newtheorem{claim}[theorem]{Claim}}
\ifndef{\result}{\newtheorem{result}[theorem]{Result}}
\ifndef{\problem}{\newtheorem{problem}{Problem}}
\ifndef{\protocol}{\newtheorem{protocol}{Protocol}}
\newtheoremstyle{myclaims}   %
{\topsep}{\topsep}   %
{\slshape}   %
{}   %
{\bfseries\itshape}   %
{}   %
{ }   %
{\thmname{#1}\thmnumber{ \!#2}.}   %
{\theoremstyle{myclaims}

\ifndef{\fact}{\newtheorem{fact}[theorem]{Fact}}
\ifndef{\observation}{\newtheorem{observation}[theorem]{Observation}}
}

\newtheoremstyle{anystatement}{\topsep}{\topsep}{\itshape}{}{\bfseries}{.}{ }{\anystatementname}
{\theoremstyle{anystatement}}

\newcommand{\anystatementname}{}

\newcounter{tmp_id_cnt}
\newcommand{\AuxNew}[4][]{#2{#3}[1][*]%
{\ifthenelse{\equal{*}{##1}} %
{\Ensuremath{#1{#4}}}%
{\ifthenelse{\equal{b}{##1}} %
{\Ensuremath{\mathbf{#4}}}%
{\ifthenelse{\equal{}{##1}} %
{\IfMathMode{#1{#4}}{#4}}{}}}}}

\newcommand{\newident}[3][*]{\ifthenelse{\equal{*}{#1}}%
{\AuxNew[\mathit]{\newcommand}{#2}{#3}} %
{\mydef{#2}{\Ensuremath{\mathit{#3}}}}} %

\newcommand{\newidentarg}[2]{%
\newcommand{#1}[1][]%
{\Ensuremath{\mathit{#2}}}}              

\newcommand{\newmat}[3][*]{\ifthenelse{\equal{*}{#1}}%
{\AuxNew{\newcommand}{#2}{#3}} %
{\mydef{#2}{\Ensuremath{#3}}}} %

\newcommand{\providemat}[3][*]{\ifthenelse{\equal{*}{#1}} %
{\AuxNew{\providecommand}{#2}{#3}} %
{\mydef{#2}{\Ensuremath{#3}}}} %

\newcommand{\providematarg}[2]{ %
\providecommand{#1}[1][]{\Ensuremath{#2}}}      

\newcommand{\newmatop}[2]{\mydef{#1}{\operatorname{#2}}}

\newcommand{\MyMakeTheoMacros}[3]{
\newcommand{#2}[2][]{\ifthenelse{\equal{}{##1}}
{\begin{#1} ##2 \end{#1}}
{\begin{#1}\label{##1} ##2\end{#1}}}
\newcommand{#3}[3][]{\ifthenelse{\equal{}{##1}}
{\begin{#1}{\e{##2}} ##3 \end{#1}}
{\begin{#1}{\e{##2}}\label{##1} ##3\end{#1}}}
}

\newcommand{\MyMakeDupTheoMacros}[8]{
\MyMakeTheoMacros{#1}{#2}{#3}
\newcommand{#4}[3]{
\newcommand{##2}{##3}
\begin{#1}\label{##1} ##2\end{#1}}
\newcommand{#5}[4]{
\newcommand{##2}{##4}
\begin{#1}{\e{##3}}\label{##1} ##2\end{#1}}
\newcommand{#8}[2]{\def\my_tmp_id{my_tmp_id_\arabic{tmp_id_cnt}}
\newtheorem*{\my_tmp_id}{#7~\ref{##1}}
\begin{\my_tmp_id} ##2 \end{\my_tmp_id}\stepcounter{tmp_id_cnt}}
\newcommand{#6}[6]{
#2[##1]{##2}

##3
\prf[#7~\ref{##1}]{##6} \newcommand{##5}{}

}
}

\newcommand{\MyMakeRefMacros}[3]{\newcommand{#1}[2][]
{\ifthenelse{\equal{}{##1}}{#2~\ref{##2}}{#3~\ref{##1} and~\ref{##2}}}}

\newcommand{\MyMakeEqRefMacros}[3]{\newcommand{#1}[2][]
{\ifthenelse{\equal{}{##1}}{#2~\eqref{##2}}{#3~\eqref{##1} and~\eqref{##2}}}}

\newcommand{\abstr}[1]{
\begin{abstract}
#1
\end{abstract}}

\newcommand{\bibentry}[8]{

\bibitem[\nospell{#8}]{#1} {\textup #3}. 
\ifthenelse{\equal{}{#6}}
{\newblock \textrm{#4.} \newblock {\em #5}, #7.}
{\newblock \textrm{#4.} \newblock {\em #5, #6}, #7.}
}

\newcommand{\inputbib}{

\bibentry{A04_Lim}{Aaronson}{S. Aaronson}{Limitations of Quantum Advice and One-Way Communication}{Proceedings of the 19th IEEE Conference on Computational Complexity}{pages 320-332}{2004}{A04}

\bibentry{BJ95_Lear_DNF}{Bshouty and Jackson}{N. Bshouty and J. Jackson}{Learning DNF over the Uniform Distribution using a Quantum Example Oracle}{Proceedings of the 8th Annual Conference on Computational Learning Theory}{pages 118-127}{1995}{BJ95}

\bibentry{BJK04_Expo}{Bar-Yossef, Jayram and Kerenidis}{Z. Bar-Yossef, T. S. Jayram and I. Kerenidis}{Exponential Separation of Quantum and Classical One-Way Communication Complexity}{Proceedings of 36th Symposium on Theory of Computing}{pages 128-137}{2004}{BJK04}

\bibentry{GKKRW07_Expo}{Gavinsky, Kempe, Kerenidis, Raz and de Wolf}{D. Gavinsky, J. Kempe, I. Kerenidis, R. Raz and R. de Wolf}{Exponential Separations for One-Way Quantum Communication Complexity, with Applications to Cryptography}{Proceedings of the 39th Symposium on Theory of Computing}{pages 516-525}{2007}{GKKRW07}

\bibentry{GKRW06_Bou}{Gavinsky, Kempe, Regev and de Wolf}{D. Gavinsky, J. Kempe, O. Regev and R. de Wolf}{Bounded-error Quantum State Identification and Exponential Separations in Communication Complexity}{Proceedings of the 38th Symposium on Theory of Computing}{pages 594-603}{2006}{GKRW06}

\bibentry{JRS02_Pri}{Jain, Radhakrishnan and Sen}{R. Jain, J. Radhakrishnan and P. Sen}{Privacy and Interaction in Quantum Communication Complexity and a Theorem about the Relative Entropy of Quantum States}{Proceedings of the 43rd Annual Symposium on Foundations of Computer Science}{pages 429-438}{2002}{JRS02}

\bibentry{KW04}{Kerenidis and de Wolf}{I. Kerenidis and R. de Wolf}{Exponential Lower Bound for 2-Query Locally Decodable Codes via a Quantum Argument}{Journal of Computer and System Sciences 69(3)}{pages 395-420}{2004}{KW04}

\bibentry{SG04_Equiv}{Servedio and Gortler}{R. Servedio and S. Gortler}{Equivalences and Separations Between Quantum and Classical Learnability}{SIAM Journal on Computing 33(5)}{pages 1067-1092}{2004}{SG04}

\bibentry{V84_A_The}{Valiant}{L. Valiant}{A Theory of the Learnable}{Communications of the ACM 27(11)}{pages 1134-1142}{1984}{V84}

}

\newcommand{\bib}[1][]{

\begin{thebibliography}{ABCDEFG00}
\frenchspacing
\inputbib
#1
\end{thebibliography}
}

\MyMakeTheoMacros{fact}{\fct}{\nfct}

\MyMakeRefMacros{\fctref}{Fact}{Facts}

\MyMakeTheoMacros{observation}{\obs}{\nobs}

\MyMakeRefMacros{\obsref}{Observation}{Observations}

\MyMakeDupTheoMacros{lemma}
{\lem}{\nlem}{\lemdup}{\nlemdup}{\lemapp}{Lemma}{\lemrep}

\MyMakeRefMacros{\lemref}{Lemma}{Lemmas}

\MyMakeDupTheoMacros{corollary}
{\crl}{\ncrl}{\crldup}{\ncrldup}{\crlapp}{Corollary}{\crlrep}

\MyMakeRefMacros{\crlref}{Corollary}{Corollaries}

\MyMakeTheoMacros{proposition}{\prp}{\nprp}

\newtheorem*{prp*}{\e{Proposition}}

\MyMakeRefMacros{\prpref}{Proposition}{Propositions}

\MyMakeDupTheoMacros{my_claim}
{\clm}{\nclm}{\clmdup}{\nclmdup}{\clmapp}{Claim}{\clmrep}

\MyMakeRefMacros{\clmref}{Claim}{Claims}

\MyMakeDupTheoMacros{theorem}
{\theo}{\ntheo}{\theodup}{\ntheodup}{\theoapp}{Theorem}{\theorep}

\MyMakeRefMacros{\theoref}{Theorem}{Theorems}

\newcommand{\faketheoref}[1]{Theorem~{#1}}

\MyMakeTheoMacros{definition}{\defi}{\ndefi}

\MyMakeRefMacros{\defiref}{Definition}{Definitions}

\MyMakeTheoMacros{problem}{\prob}{\nprob}

\MyMakeRefMacros{\probref}{Problem}{Problems}

\MyMakeTheoMacros{protocol}{\prot}{\nprot}

\MyMakeRefMacros{\protref}{Protocol}{Protocols}

\providecommand{\qedsymbol}{\square}

\newcommand{\prf}[2][]{\ifthenelse{\equal{}{#1}}%
{\begin{proof}\renewcommand{\qedsymbol}{$\blacksquare$}%
#2 \end{proof}}%
{\begin{proof}[Proof of #1]%
\renewcommand{\qedsymbol}{$\blacksquare_{\mbox{\it{\scriptsize{#1}}}}$}%
#2 \end{proof}}
}

\newcommand{\sect}[2][]{\ifthenelse{\equal{}{#1}}
{\section{#2}}
{\section{#2}\label{#1}}}

\newcommand{\ssect}[2][]{\ifthenelse{\equal{}{#1}}
{\subsection{#2}}
{\subsection{#2}\label{#1}}}

\MyMakeRefMacros{\chref}{Chapter}{Chapters}

\MyMakeRefMacros{\sref}{Section}{Sections}

\MyMakeRefMacros{\ssref}{Subsection}{Subsections}

\MyMakeRefMacros{\sssref}{Subsection}{Subsections}

\definecolor{DarkGreen}{rgb}{0,0.45,0.08}
\definecolor{LightBlue}{rgb}{0.122,0.016,0.855}

\newcommand{\IfMathMode}[2]{\ifmmode{#1}\else{#2}\fi}

\newcommand{\Ensuremath}{\ensuremath}

\newcommand{\fbr}[1]{\IfMathMode %
{#1}{$#1$}}                      %

\newcommand{\fnbr}[1]{\mbox{\fbr{#1}}}   %

\newcommand{\fla}[2][*]{\ifthenelse{\equal{}{#1}}{\fbr{#2}}{\fnbr{#2}}}

\newcommand{\mat}[2][]{\ifthenelse{\equal{}{#1}} %
{ \begin{displaymath} #2 \end{displaymath} } %
{ \begin{equation} \label{#1} #2 \end{equation} }%
}

\newcommand{\matal}[2][]{\mat[#1]{\begin{aligned} #2 \end{aligned}}}

\newcommand{\f}{\fla}

\newcommand{\m}{\mat}
\newcommand{\mal}{\matal}

\MyMakeEqRefMacros{\equref}{Equation}{Equations}

\MyMakeEqRefMacros{\expref}{Expression}{Expressions}

\MyMakeEqRefMacros{\inequref}{Inequality}{Inequalities}

\newcommand{\bracref}[1]{(\ref{#1})}

\newcommand{\bref}{\bracref}

\MyMakeRefMacros{\figref}{Figure}{Figures}

\providecommand{\middle}{\big}

\newmatop{\poly}{poly}

\newcommand{\h}[2][]{\ifthenelse{\equal{}{#2}}%
{\mathop{\mathbf{H}}_{#1}}%
{\mathop{\mathbf{H}}_{#1}{\left[{#2}\right]}}}
\newcommand{\hh}[3][]{\mathop{\mathbf{H}}_{#1}%
{\left[{#2}\middle|\vphantom{|_1^1}{#3}\right]}}

\newcommand{\KL}[2]{d_{KL}\llp{#1}\middle|\middle|{#2}\rrp}

\newcommand{\I}[3][]{\ifthenelse{\equal{}{#1}}%
{\mathbf{I}{\left[{#2}:{#3}\right]}}%
{\mathbf{I}{\left[{#2}:{#3}\middle|{#1}\right]}}}

\providecommand{\E}[2][]{\mathop{\mathbf{E}}_{#1}{\left[{#2}\right]}}

\newcommand{\PR}[2][]{\mathop{\mathbf{Pr}}_{#1}{\left[{#2}\right]}}

\newcommand{\pss}[1][]{\nospell{\ifthenelse{\equal{}{#1}}%
{\txt{'s}}%
{\fla{#1\txt{'s}}}}}

\newcommand{\ord}[1][]{\nospell{\ifthenelse{\equal{}{#1}}%
{\txt{'th}}%
{\ifthenelse{\equal{1}{#1}}{$1\txt{'st}$}{\ifthenelse{\equal{2}{#1}}{$2\txt{'nd}$}{\ifthenelse{\equal{3}{#1}}{$3\txt{'rd}$}{\fla{#1\txt{'th}}}}}}}}

\newmat[]{\llp}{\left(}
\newmat[]{\rrp}{\right)}

\newmat[]{\rra}{\right\rangle}

\newmat[]{\llc}{\left\lceil}
\newmat[]{\rrc}{\right\rceil}

\newmat[]{\dt}{\cdot}
\newmat[]{\tm}{\cdot}
\newmat[]{\xor}{\oplus}
\newmat[]{\sbseq}{\subseteq}
\newmat[]{\nsbse}{\nsubseteq}
\newmat[]{\sbs}{\subset}
\newmat[]{\smin}{\setminus}
\newmat[]{\eps}{\varepsilon}
\newmat[]{\deq}{\stackrel{\textrm{def}}{=}}

\providemat{\QQ}{\mathbb{Q}}
\providematarg{\NN}{\ifthenelse{\equal{}{#1}}%
{\mathbb{N}}%
{\mathbb{N}_{#1}}}
\providematarg{\RR}{\ifthenelse{\equal{}{#1}}%
{\mathbb{R}}%
{\mathbb{R}^{#1}}}
\providematarg{\ZZ}{\ifthenelse{\equal{}{#1}}%
{\mathbb{Z}}%
{\mathbb{Z}_{#1}}}

\newmat[]{\ds}{\dots}
\newmat[]{\dc}{,\ds,}

\newcommand{\itemi}[2][]{\ifthenelse{\equal{}{#1}}
{\begin{itemize} #2 \end{itemize}}
{\begin{itemize}[#1] #2 \end{itemize}}}

\newcommand{\wrt}{w.r.t.\ }	 %

\newcommand{\fr}[3][*]{%
\ifthenelse{\equal{*}{#1}}        %
{\frac{#2}{#3}}{}%
\ifthenelse{\equal{/}{#1}}        %
{\nicefrac{#2}{#3}}{}%
\ifthenelse{\equal{}{#1}}         %
{\left.#2\middle/#3\right.}{}%
\ifthenelse{\equal{p_}{#1}}       %
{\left.\left(#2\right)\middle/#3\right.}{}%
\ifthenelse{\equal{_p}{#1}}       %
{\left.#2\middle/\left(#3\right)\right.}{}%
\ifthenelse{\equal{pp}{#1}}       %
{\left.\left(#2\right)\middle/\left(#3\right)\right.}{}
}

\def\MySQRT#1#2{    %
\setbox0=\hbox{$#1\sqrt{#2\,}$}\dimen0=\ht0%
\advance\dimen0-0.2\ht0%
\setbox2=\hbox{\vrule height\ht0 depth -\dimen0}%
{\box0\lower0.4pt\box2}}

\newcommand{\set}[2][]{\ifthenelse{\equal{}{#1}} %
{\Ensuremath{\left\{#2\right\}}}%
{\Ensuremath{\left\{#2\,\middle\arrowvert\,#1\right\}}}}

\newcommand{\Sup}[2][]{\ifthenelse{\equal{}{#1}} %
{\Ensuremath{\sup{\left\{#2\right\}}}}%
{\Ensuremath{\sup{\left\{#2\,\middle\arrowvert\,#1\right\}}}}}

\newcommand{\newfunction}[2]{ %
\newcommand{#1}[2][*]{\ifthenelse{\equal{*}{##1}}%
{\Ensuremath{#2{\left(##2\right)}}}%
{#2(##2)}}%
}

\newfunction{\asO}{O}
\newfunction{\aso}{o}
\newfunction{\asOm}{\Omega}

\providecommand{\ket}[1]{\Ensuremath{\left|#1\rra}}
\newmat{\kI}{\ket{1}}
\newmat{\kO}{\ket{0}}

\newcommand{\kbra}[2][]{\ifthenelse{\equal{}{#1}}%
{\Ensuremath{\left|#2\middle\rangle\hspace{-2.5pt}\middle\langle #2\right|}}%
{\Ensuremath{\left|#1\middle\rangle\hspace{-2.5pt}\middle\langle #2\right|}}%
}

\mydef{\01}{\set{0,1}}

\renewcommand{\l}{\left}
\renewcommand{\r}{\right}

\newcommand{\sz}[2][]{\ifthenelse{\equal{}{#1}}%
{\Ensuremath{\left|#2\right|}}%
{\Ensuremath{\left|#2\middle|_{#1}\right.}}}

\providecommand{\ceil}[2][*]{\ifthenelse{\equal{}{#1}}%
{\lceil #2 \rceil}%
{\llc #2 \rrc}}

\newcommand{\fn}{\footnote}

\newcommand{\nin}{\not\in}   %

\newcommand{\e}{\emph}

\newcommand{\ul}[1]{\underline{#1}}  %

\newcommand{\txt}[1]{\textrm{#1}}   %

\newcommand{\Cl}[1]{{\cal #1}} %

\DeclareMathAlphabet{\lowcal}{OT1}{pzc}{m}{it}

\newidentarg{\RI}{\mathcal R_{#1}^{1}}
\newidentarg{\QI}{\mathcal Q_{#1}^{1}}

\newident[]{\P}{P}
\newident{\BQP}{BQP}

\usepackage{hyphenat}   %

\date{}
\newident{\qac}{QAC}
\newident{\pq}{PQ}
\newident{\pac}{PAC}

\title{Quantum Predictive Learning and\\
Communication Complexity with Single Input}

\author{
{\bf Dmitry Gavinsky}\\
{\small NEC Laboratories America, Inc.}\\
{\small 4 Independence Way, Suite 200}\\
{\small Princeton, NJ 08540, U.S.A.}
}

\begin{document}

\maketitle

\thispagestyle{empty}

\abstr{We define a new model of quantum learning that we call \e{Predictive Quantum (\pq)}.
This is a quantum analogue of \pac, where during the testing phase the student is only required to answer a polynomial number of testing queries.

We demonstrate a relational concept class that is \e{efficiently learnable} in \pq, while in \e{any} ``reasonable'' classical model exponential amount of training data would be required.
This is the first unconditional separation between quantum and classical learning.

We show that our separation is the best possible in several ways; in particular, there is no analogous result for a functional class, as well as for several weaker versions of quantum learning.

In order to demonstrate tightness of our separation we consider a special case of one-way communication that we call \e{single-input mode}, where Bob receives no input.
Somewhat surprisingly, this setting becomes nontrivial when relational communication tasks are considered.
In particular, any problem with two-sided input can be transformed into a single-input relational problem of equal \e{classical} one-way cost.
We show that the situation is different in the \e{quantum} case, where the same transformation can make the communication complexity exponentially larger.
This happens if and only if the original problem has exponential gap between quantum and classical one-way communication costs.
We believe that these auxiliary results might be of independent interest.}

\setcounter{page}{1}

\sect{Introduction}

In this paper we compare quantum and classical modes of computational learning and give the first unconditional exponential separation between the two.

Let $X$ be a (finite) domain and $Y$ be a set of possible labels.
Let $\Cl C$ be a \e{concept class} consisting of functions $\ell:X\to Y$, each $\ell\in\Cl C$ can be viewed as assignment of a label to every $x\in X$.
The knowledge of $X$, $Y$ and $\Cl C$ is shared between a \e{teacher} and a \e{learner}; the teacher also knows some \e{target concept} $\ell_0\in\Cl C$, unknown to the learner.
The learning process consists of two stages: the \e{learning phase}, followed by the \e{testing phase}.
In the learning phase, the teacher and the learner communicate in order to let the latter learn $\ell_0$.
In the testing phase, the learner has to demonstrate that he has successfully learned $\ell_0$: for example, an arbitrary $x\in X$ may be given to him, and he would have to respond with $\ell_0(x)$.

A \e{learning model} specifies the set of rules governing the learning and the testing phases.
The teacher is, in general, viewed as an adversary that obeys the model's restrictions.

One of the most natural and widely used learning models is that of \e{Probably Approximately Correct (\pac)}, defined by Valiant~\cite{V84_A_The}.
In the learning phase of \pac\ a sequence of training examples
\m{(x_1,\ell_0(x_1))\dc(x_k,\ell_0(x_k))}
is sent by the teacher to the learner.
The examples are independently chosen according to some distribution $D$ over the domain $X$.\fn
{Several variations of \pac\ are studied in the literature, in particular there is a definition that allows ``distribution-specific'' learning algorithms.
In this paper we will always fix $D$ to be the uniform distribution over $X$, as that is sufficient for our purposes and simplifies the notation at the same time.}
In the testing phase the learner is given a random $x\sim D$ and has to respond with $\ell_0(x)$.

Two error parameters are present in the definition of \pac:\ \e{accuracy} $1-\eps$ and \e{confidence} $1-\delta$.
We say that learning was successful if in the testing phase the learner correctly labels a randomly chosen $x\sim D$ with probability at least $1-\eps$.
A learning algorithm must be successful with probability at least $1-\delta$, taken over both algorithm's randomness and the set of examples received during the learning phase.

We say that a concept class $\Cl C$ is \e{efficiently learnable} in \pac\ if there exists an algorithm that runs in time at most polylogarithmic in the domain size and polynomial in $1/\eps$ and $1/\delta$, and learns any $\ell\in\Cl C$.
Note that the running time of an algorithm is, trivially, an upper bound on the number of training examples that it uses during the learning phase.

\ssect{Previous work}

In \cite{BJ95_Lear_DNF} Bshouty and Jackson introduced a natural quantum analogue of \pac, which we denote here by \qac.
They gave an efficient algorithm that learns DNF formulas \wrt the uniform distribution from \e{quantum} examples -- this is currently not known to be possible from classical examples (even with a quantum learning algorithm).

The question of whether quantum learning models are more efficient than the classical ones has been considered by Servedio and Gortler~\cite{SG04_Equiv}, who showed that the models \pac\ and \qac\ are equivalent from the information-theoretic point of view.
On the other hand, they showed that quantum models are computationally more powerful than their classical analogues if certain cryptographic assumptions hold.

\ssect{Our results}

In the definition of a new learning model \pq\ (\e{Predictive Quantum}) we will generalize \qac\ in several ways.

First, we allow \e{relational} concept classes.
Namely, the elements $\ell$ of $\Cl C$ can be arbitrary subsets of $X\times Y$, thus allowing multiple correct labeling for every $x\in X$.
During the learning phase the learner receives pairs $(x_i,y_i)$, such that $x_i\sim D$ and $y_i$ is a uniformly random element of \set[(x_i,y)\in\ell_0]{y}.
At the testing phase any $y$ satisfying $(x,y)\in\ell_0$ is accepted as a correct answer to the query $x$. 

Second, we classify all learning models as follows:
\itemi{
\item We call \e{standard} a learning model where in the testing phase the learner outputs a \e{final hypothesis}, viewed as a function $h:X\to Y$.
In the testing phase it is checked whether $h(x)$ agrees well with the target concept.
The final hypothesis should be efficiently evaluatable (under the same notion of efficiency that applies to the learning algorithms in the model).
\item We say that a model is \e{quasi-predictive} if the learner has to answer queries in the testing phase.
The number of testing queries that will be asked is unknown during the learning phase.
\item We call a model \e{predictive} if the learner should answer a single query in the testing phase.\fn
{Note that a concept class that is efficiently learnable by our definition of predictive learning is also efficiently learnable in a version where \e{polynomial number} of testing queries are made.
For notational convenience we will use the single-query definition of predictive learnability.}
}

For example, the \pac\ model, as defined above, is predictive.
If we would allow an arbitrary number of testing queries, that would make it quasi-predictive.
If we require that in the end of the learning phase the learner produces a hypothesis $h:X\to Y$, such that $\PR[x\sim D]{h(x)=\ell(x)}\ge1-\eps$, that turns the model into standard.

As long as the learning phase remains unchanged, standard learnability of a concept implies its quasi-predictive learnability, which, in turn, implies predictive learnability.
On the other hand, it is well known that in any ``reasonable'' \e{classical} learning model, a predictive learning algorithm can be turned into a standard one (this can be achieved by producing a final hypothesis consisting of a description of the answering subroutine, all the data available after the learning phase, and a random string, if randomness is used by the answering subroutine).
Therefore, in the classical case the standard, the quasi-predictive, and the predictive modes of learning are essentially equivalent; in particular, the above three definitions of \pac\ give rise to the same family of efficiently learnable concept classes.
We will see that the situation is different with quantum learning.

For the rest of the paper let $n\deq\ceil{\log\sz X}$.
Consider the following definition.
\defi[d_appr]{Let $D$ be a distribution over $X$.
We say that a hypothesis $h:X\to Y$ approximates a concept $\ell\in\Cl C$ \wrt $D$ if
\itemi{
\item $\PR[x\sim D]{h(x)=\ell(x)}\ge2/3$, when $\ell:X\to Y$ is a function;
\item $\PR[x\sim D]{\llp x,h(x)\rrp\in\ell}\ge2/3$, when $\ell\sbseq X\times Y$ is a relation.
}
A hypotheses class $\Cl H$ is said to approximate $\Cl C$ if for every $\ell\in\Cl C$, $\Cl H$ contains some $h$ that approximates $\ell$.
}

Any standard algorithm that learns $\Cl C$ with $\eps\le1/3$ must use a class of final hypotheses that approximates $\Cl C$.
An efficient algorithm can use a class of final hypotheses of size at most exponential in $\poly(n)$.
As outlined above, efficient learnability in any classical model implies efficient learnability in the corresponding standard model, and therefore \e{$\Cl C$ is efficiently learnable in some classical model only if there exists $\Cl H$ of size at most $2^{\poly(n)}$ that approximates $\Cl C$}.

We call a concept class $\Cl C$ \e{unspeakable} if any class $\Cl H$ that approximates it should be of size at least $2^{2^{\asOm n}}$.
In particular, \e{neither a classical algorithm nor a standard quantum algorithm can efficiently learn an unspeakable concept class}.

In this paper we demonstrate an \e{efficient quantum predictive algorithm that learns an unspeakable relational concept class}.
Therefore, \e{quantum predictive learnability does not imply quantum standard learnability}.
On the other hand, we will show that \e{no quasi-predictive quantum algorithm can efficiently learn an unspeakable concept class}.
We also show that \e{efficient quantum learning of a functional unspeakable concept class is impossible}, and therefore the combination of relational concepts and quantum predictive mode of learning is essential for learning an unspeakable class.

Following is a summary of our main results (cf.\ \theoref{t_main}, \lemref{l_no_func}, and \lemref{l_no_quasi}).
\theo[t_+]{There exists a relational concept class that is unspeakable but can be efficiently learned in the model of predictive quantum \pac.
}

A concept class $\Cl C$ that witnesses the above theorem is given in \defiref{d_C}.
Its construction has been inspired by a communication problem due to Bar-Yossef, Jayram and Kerenidis~\cite{BJK04_Expo}.

\theo[t_-]{Classical learning of an unspeakable concept class is not possible from less than exponential amount of information from the teacher, even by a computationally unlimited learner.

Both standard and quasi-predictive learning of an unspeakable concept class is not possible from less than exponential amount of quantum information from the teacher, even by a computationally unlimited learner.

Predictive learning of an unspeakable functional concept class is not possible from less than exponential amount of quantum information from the teacher, even by a computationally unlimited learner.
}

Two parts of \theoref{t_-} are proved by making connection to two ``impossibility of separation'' results in communication complexity.
One of them is due to Aaronson~\cite{A04_Lim}, and the other is new and might be of independent interest.

We will consider a special case of one-way communication, which will we call \e{single-input mode}, where Bob receives no input.
We show that, somewhat surprisingly, for any single-input communication task the quantum and the classical one-way costs are asymptotically the same (the statement is trivial for functional tasks, but the relational case is more involved).
More details can be found in \sref{ss_single}.

\sect[s_defi]{Definitions and more}

For $a\in\NN$ we denote $[a]\deq\set{1\dc a}$.
We view the elements of $\ZZ[a]$ as integers \set{0,1,\dc a-1}, and accordingly we define their ordering $0<1<\ds<a-1$.
For any $i\in\NN$ and $b\in\ZZ[a]$, let $i\tm b=ib$ be the \ord[i] power of $b$ \wrt the group operation $+$.
We use subscripts to address individual bits of binary strings: for $x\in\01^n$ and $i\in[n]$, $x_i$ stands for the \ord[i] bit of $x$.

Let $\KL\dt\dt$ denote the \e{relative entropy}, or the \e{Kullback-Leibler divergence}.
We will need the following \e{substate lemma}, due to Jain, Radhakrishnan and Sen~\cite{JRS02_Pri}.

\nlem[l_substate]{\cite{JRS02_Pri}}{For any distributions $\mu_1$ and $\mu_2$ over the finite sample space $X$ and $r\ge1$,
\m{\PR[x\sim\mu_1]
{\fr{\mu_1(x)}{\mu_2(x)}>2^{r\l(\KL{\mu_1}{\mu_2}+1\r)}}<\fr1r.}}

For completeness, below we give its proof.

\prf{Let \f{X'\deq\set[{\fr[/]{\mu_1(x)}{\mu_2(x)}\le2^{r\l(\KL{\mu_1}{\mu_2}+1\r)}}]{x\in X}.}
Then by concavity of the logarithm function,
\m{\mu_1(X')\log\fr{\mu_1(X')}{\mu_2(X')}
+\mu_1(X\smin X')\log\fr{\mu_1(X\smin X')}{\mu_2(X\smin X')}
\le\KL{\mu_1}{\mu_2}.}
On the other hand,
$\mu_1(X')\log\fr{\mu_1(X')}{\mu_2(X')}\ge\inf_{x\in(0,1]}x\log x>-1$, and
\m{\PR[x\sim\mu_1]{x\nin X'}\tm r\l(\KL{\mu_1}{\mu_2}+1\r)<
\KL{\mu_1}{\mu_2}+1,}
as required.}

Let $D$ be the uniform distribution over $X$, recall \defiref{d_appr}.
\defi[d_unsp]{Let $\Cl C$ be a concept class.
We say that $\Cl C$ is unspeakable if $\sz{\Cl C'}\in2^{2^{\asOm n}}$ holds for any $\Cl C'$ that approximates $\Cl C$ \wrt $D$.}

\ssect{Quantum learning}

In \cite{SG04_Equiv} the authors provide an excellent survey of quantum vs.\ classical learning.
Below we sketch one possible intuition behind the concepts considered in this work.\fn
{This part is mostly meant to assist a reader whose familiarity with quantum computing is limited; analyzing the philosophical foundations of quantum mechanics is beyond our scope.}

Starting from \pac, how can we make it quantum?
First, we can give the student ability to run any computation that a quantum computer can perform efficiently (e.g., to decide membership in any language from the complexity class \BQP).
Second, we can let the training examples be quantum, i.e., the student receives from the teacher \e{quantum bits (qubits)}.
In this paper we consider the situation when \e{both the student and the examples are quantum}.

Information-theoretic consequences of ``quantumness'' stem from the facts that, on the one hand, quantum states require \e{exponential} (in the number of qubits) amount of classical bits for their full description, while on the other hand, the uncertainty principle dictates that given a quantum state only a (tiny) fraction of that classical data can be accessed by an observer.

Note also that computational impact of a student being quantum is not necessarily captured by the power of \BQP:
As training examples are quantum, the student can apply quantum algorithms to quantum input, while \BQP\ only deals with situations when quantum algorithms are fed with classical input.

What can be viewed as a reasonable model of quantum training examples?
Let the target concept be $\ell_0$.
First, assume that $\ell_0:X\to Y$ is a Boolean function, then a quantum example shall look like
\m{\fr1{\sqrt{|X|}}\sum_{x\in X}\ket{x,\ell_0(x)},}
where $\ket{\dt}$ denotes the corresponding basis state
of the quantum register over $n+1$ qubits.
Note that the above form of training examples corresponds to the uniform distribution of $x\in X$, since measuring the first $n$ qubits in the computational basis can return each possible $x_0\in X$ with the same probability of $1/|X|$.

Now, let $\ell_0\sbseq X\times Y$ be a relation.
We need a quantum superposition over all possible pairs $(x,y)\in\ell_0$.
Naturally, we want to choose the amplitudes such that every $x_0$ still shows up with probability $1/|X|$, and at the same time, conditional on obtaining $x_0'$, every element of $\set[(x_0',y)\in\ell_0]{y}$ appears with equal probability. 
It can be seen that the following quantum superposition satisfies the requirements:
\m{\sum_{(x,y)\in\ell_0}
\fr1{\sqrt{\sz X\tm\sz{\set[(x,y')\in\ell_0]{y'}}}}\ket{x,y}.}
This quantum state will be used in \defiref{d_qacs} below to describe the training examples that our student will receive from the teacher.

\ssect{The model of predictive quantum learning}

We will usually ignore normalization factors and global phases of quantum states.\fn
{That is, we allow arbitrary non-zero complex vectors to represent quantum states.}
We define a predictive quantum version of \pac, as follows.

\defi[d_qacs]{In the \pq\ (Predictive Quantum) learning model, a learning algorithm can ask for arbitrarily many copies of the state 
\m{\sum_{(x,y)\in\ell_0}\fr1{\sqrt{\sz{\set[(x,y')\in\ell_0]{y'}}}}\ket{x,y},}
where $\ell_0\sbseq X\times Y$ is a relational target concept.
In the end of the learning process the algorithm receives an element $x\in X$ and should, with probability at least $5/6$, output any $y$ satisfying $(x,y)\in\ell_0$.

A learning algorithm is efficient if its running time is at most polynomial in $n\deq\ceil{\log|X|}$.
A concept class $\Cl C$ is efficiently learnable in \pq\ if there exists an efficient algorithm that \pq-learns every $\ell\in\Cl C$.}

In the above definition the relative amplitudes of the pairs $\ket{x,y}$ in a training example are chosen such that a projective measurement in the computational basis would result in a uniformly chosen $x$, and given $x$, all elements of $\set[(x,y')\in\ell_0]{y'}$ are equally likely to come with it.
Therefore, the model can be viewed as a natural quantum generalization of the relational version of \pac, as discussed in the Introduction.

The fact that all quantum training examples are the same lets us get rid of the confidence parameter ($\delta$) in the definition of \pq\ (there is no such thing as ``unlucky'' sample of training examples).
For simplicity, we choose the required accuracy ($\eps$) to always be $5/6$.
Note also that in the testing phase we want the learning algorithm to give a correct answer to \e{any} $x\in X$ with good probability (instead of just being able to cope with a randomly chosen $x$).
This further simplifies the definition and also makes our result stronger (as we construct a \pq-algorithm, and do not state any lower bound against this model).

\sect[s_C]{Concept class $\Cl C$}

We define a concept class $\Cl C$ that will be shown to be both unspeakable and efficiently \pq-learnable.
Our definition has been inspired by a communication problem considered in \cite{BJK04_Expo}.
\defi[d_C]{Let $N$ be prime.
Every concept in the class $\Cl C$ is represented by $C\in\01^N$.
The set of queries is $[N-1]$, represented by binary strings of length $n=\ceil{\log N}$.
A pair $(x,b)\in\ZZ[N]\times\01$ is a valid answer to query $j$ \wrt $C\in\Cl C$ if $C_x\xor C_{x+j}=b$.}

We slightly abuse the notation by viewing each $C\in\Cl C$ either as a binary string of length $N$ or as a set \set[\txt{$(x,b)$ is a valid answer to $j$ \wrt $C$}]{(j,x,b)}.

\theo[t_main]{The concept class $\Cl C$ is unspeakable.
On the other hand, $\Cl C$ is efficiently learnable in \pq.}

The two parts of the theorem will be proved in \sref[ss_effi]{ss_unspeak}, respectively.
The key observation that we use to efficiently learn $\Cl C$ is the following (originating from \cite{KW04}).
Let a binary string $x\in\01^n$ be represented as a quantum state $\ket{\alpha(x)}=\sum(-1)^{x_i}\ket i$, where $i$ ranges in $[n]$.
Even though it is impossible to recover individual bits of $x$ by measuring $\ket{\alpha(x)}$, there is something nontrivial about $x$ that can be learned from $\ket{\alpha(x)}$.
Namely, given any perfect matching $M$ over $[n]$, it is possible to measure $\ket{\alpha(x)}$ in such a way that for some $(i,j)\in M$ the value of $x_i\xor x_j$ would become known after the measurement.
The quantum state $\ket{\alpha(x)}$ fits in $\ceil{\log n}$ qubits; on the other hand, it can be shown that the amount of classical information needed to allow similar type of access to $x$ is $n^{\asOm1}$, and this is used to show that $\Cl C$ is unspeakable.

\ssect[ss_effi]{Efficient \pq-learning of $\Cl C$}

Our learner will need $k$ \pq-examples in order to answer to the testing query with probability $1-1/2^k$, and whenever an answer is given it is correct.\fn
{If we allow a slightly modified form of training examples, where $i$ is represented through the amplitude as $\sum_{(j,x,i)\in C}(-1)^i\ket{j,x}$, then it is possible to \pq-learn $\Cl C$ \e{exactly} from one such example.}
Fix $C\in\Cl C$, then the training examples are of the form
\m{\ket{\alpha^C}\deq\sum_{(j,x,i)\in C}\ket{j,x,i}.}
The learner measures the last register of each of the $k$ instances of $\ket{\alpha^C}$ in the basis $\{\kO+\kI, \kO-\kI\}$.
With probability $1-1/2^k$ at least one measurement results in $\kO-\kI$, then the learner keeps that copy and abandons the rest (otherwise he gives up).
Next, the learner measures the second register in the computational basis, thus obtaining in the first two registers
\m{\sum_{(j,x_0,i)\in C}(-1)^i\ket{j,x_0}=
\sum_{j\in[N-1]}(-1)^{C_{x_0}\xor C_{x_0+j}}\ket{j,x_0}=
\sum_{j\in[N-1]}(-1)^{C_{x_0+j}}\ket{j,x_0}}
for some $x_0\in\ZZ[N]$.
Then he performs the transformation $\ket{j,x_0}\to\ket{j+x_0,x_0}$, and the state of the first register becomes
\m{\ket{\alpha_{x_0}^C}\deq
\sum_{j\in[N-1]}(-1)^{C_{x_0+j}}\ket{x_0+j}=
\sum_{k\in\ZZ[N]\smin\set{x_0}}(-1)^{C_k}\ket{k}.}

At this point the learner is ready for the testing phase.
Assume that a question $q\in[N-1]$ has been asked.
Define the following perfect matching over $\ZZ[N]\smin\set{x_0}$:
\m{m_q\deq\set[0\le i\le\fr{N-3}2]{\big(x_0+(2i+1)q, x_0+(2i+2)q\big)}.}
Pairwise disjointness of the edges and the fact that $x_0$ is isolated follow from primality of $N$.
The learner performs projective measurement of $\ket{\alpha_{x_0}^C}$ onto $(N-1)/2$ sub-spaces, each spanned by a pair of vectors $\ket a$ and $\ket b$ where $a$ and $b$ are connected in $m_q$ (to make the measurement complete we add $\kbra{x_0}$ to it, but this outcome never occurs).

Assume that the outcome of the last measurement corresponds to the edge $(a,a+q)\in m_q$.
Then the state of the register that contained $\ket{\alpha_{x_0}^C}$ becomes either $\ket a+\ket{a+q}$ or $\ket a-\ket{a+q}$, the former corresponding to $C_a\xor C_{a+q}=0$ and the latter to $C_a\xor C_{a+q}=1$.
As the two states are orthogonal, the learner is able to distinguish and, respectively, answer $(a,0)$ in the first case and $(a,1)$ in the second, and that is a correct answer.

All quantum operations involved in the algorithm can be performed efficiently.

\ssect[ss_unspeak]{$\Cl C$ is unspeakable}

Let us see that the concept class $\Cl C$ is unspeakable.
The following proof uses some ideas from \cite{BJK04_Expo} and~\cite{GKRW06_Bou}.

Assume that $\Cl C$ is approximated by a class $\Cl D$.
Then there exists some $h_0\in\Cl D$ that simultaneously approximates at least $2^N\big/\sz{\Cl D}$ elements of $\Cl C$, denote the set of those elements by $\Cl C_0$.

Consider the answers
that $h_0$ gives to all possible queries $q\in[N-1]$.
Denote $(x_q,i_q)\deq h_0(q)$ and let
\m{Q_0\deq\set[\txt{$(x_q,i_q)$ is a good answer to $q$ \wrt at least \ord[3/5] of \pss[\Cl C_0] elements}]{q}.}
Counting reveals that $\sz{Q_0}\ge\fr{N-1}6$.

Let $e_q\deq(x_q,x_q+q)$ and $E_0\deq\set[q\in Q_0]{e_q}$.
Every edge $e_q$ corresponds to at most $2$ different values of $q\in[N-1]$,
therefore $\sz{E_0}\ge\fr{N-1}{12}$.
Consider a graph $G_0$ over $N$ nodes, whose edges are the elements of $E_0$.
Observe that $G_0$ contains at least $\sqrt{2\sz{E_0}}\ge\sqrt{\fr{N-1}6}$ non-isolated vertices.

Let $F_0\sbseq G_0$ be a forest consisting of a spanning tree for each connected component of $G_0$.
Then $F_0$ contains at least $\sqrt{\fr{N-1}{24}}$ edges, denote them by $E_0'$.
Let $Q_0'\sbseq Q_0$ be a subset of size $\sz{E_0'}$, such that
\m{E_0'=\set[q\in Q_0']{e_q}.}

View the elements of $\Cl C$ as binary strings of length $N$.
Let us consider two probability distributions, one corresponding to uniformly choosing $C\in\Cl C$ and the other corresponding to uniformly choosing $C\in\Cl C_0$ -- denote them by $D^C$ and $D_0^C$, respectively.
Then
\m{\log\llp\fr{\sz{\Cl C}}{\sz{\Cl C_0}}\rrp = \h{D^C}-\h{D_0^C},}
where $\h{\dt}$ denotes the binary entropy.

For every $e_q=(a,b)$ put $I_q\deq C_a\xor C_b$, and let $J\deq(I_q)_{q\in Q_0'}$.
It is straightforward from the construction of $Q_0'$ that if $C\sim D^C$ then the collection \set[q\in Q_0']{I_q} consists of mutually independent unbiased Boolean random variables, and therefore $\h[D^C]{J}=\sz{Q_0'}$.

As \f{\h{C}=\h{J}+\hh{C}{J}} holds \wrt any distribution of $C$,
\mal[m_ent]{\log\llp\fr{\sz{\Cl C}}{\sz{\Cl C_0}}\rrp
&=\h[D^C]{C}-\h[D_0^C]{C}
=\h[D^C]{J}-\h[D_0^C]{J}+\hh[D^C]{C}{J}-\hh[D_0^C]{C}{J}\\
&\ge\h[D^C]{J}-\h[D_0^C]{J}=\sz{Q_0'}-\h[D_0^C]{J}\\
&\ge\sz{Q_0'}-\sum_{q\in Q_0'}\h[D_0^C]{I_q}
=\sum_{q\in Q_0'}\llp1-\h[D_0^C]{I_q}\rrp,}
where the first inequality follows from the fact that $\hh[D^C]{C}{J}=N-\sz{Q_0'}$, which is the maximum of $\hh{C}{J}$ under any distribution of $C$.

From the definition of $Q_0$ (and the fact that $Q_0'\sbseq Q_0$), we know that each of \set[q\in Q_0']{I_q} is at least $3/5$-biased, therefore $\h[D_0^C]{I_q}\le\fr{49}{50}$, and \bref{m_ent} leads to
\m{\log\llp\fr{\sz{\Cl C}}{\sz{\Cl C_0}}\rrp
\ge\fr{\sz{Q_0'}}{50}=\fr{\sz{E_0'}}{50}>\fr{\sqrt N}{250},}
for sufficiently large $N$.
According to our choice of $h_0$,
\m{\sz{\Cl D}\ge\fr{\sz{\Cl C}}{\sz{\Cl C_0}}
\in2^{N^{\asOm1}}\sbseq2^{2^{\asOm n}},}
which means that the class $\Cl C$ is unspeakable.

\sect{Optimality of our separation}

The model of \pq\ where we demonstrated learnability of $\Cl C$ is computationally feasible.
But in the definition of \pq\ we have modified what is probably the most usual learning setting in several ways:
Besides being quantum, our algorithm is \e{predictive}; moreover, the concept class that we learn is a \e{relational} one.
In this section we will see that all these ``enhancements'' are essential in order to be able to learn an unspeakable class efficiently.

We already know that classical learning of an unspeakable class cannot be efficient.
We will show that exponential amount of training data is required in order to learn a functional unspeakable concept (\lemref{l_no_func}), as well as to learn any unspeakable concept in quasi-predictive setting (\lemref{l_no_quasi}).
The both results are established through making a connection to one-way communication complexity:
Our proof of \lemref{l_no_func} is based on Aaronson's~\cite{A04_Lim}, and in order to prove \lemref{l_no_quasi} we establish a new fact about one-way communication complexity that might be of independent interest (\theoref{t_QR}, \crlref{cr_QR}).

\ssect{Quantum and classical one-way communication complexity}

The one-way model of communication complexity is defined as follows.
Let $P\sbseq X\times Y\times Z$ be a (relational) two-party communication problem.
Input to $P$ has the form $(x,y)\in X\times Y$, in the beginning it is split between two players: Alice receives $x$ and Bob receives $y$.
The goal is for Bob to produce $z\in Z$, such that $(x,y,z)\in P$.
The players cooperate to achieve it, namely Alice sends a message $m$ to Bob, and he outputs $z\in Z$ based on the message $m$ and his portion of input $y$.

Assume for convenience that both the length of $y$ and the length of $m$ are functions of the lengths of $x$, and denote the latter by $n=\ceil{\log|X|}$.
Both Alice and Bob are all-powerful computationally, and their goal is to solve the problem using as short an $m$ as possible.
There are two versions of this model that we are interested in, namely \e{quantum} and \e{classical}.
In the former the action of the players should obey the laws of quantum mechanics, in particular the message $m$ is quantum and its ``length'' is measured in qubits; in the latter the message is classical and consists of bits.
We let our protocols employ mixed strategies, i.e., shared randomness is allowed.

For any $\eps$ we say that \e{a protocol $\cal T$ solves $P$ with error $\eps$} if Alice and Bob, who behave according to $\cal T$, produce a correct answer to every input $(x,y)\in X\times Y$ with probability at least $1-\eps$.
For a distribution $\mu$ over $X\times Y$ we say that \e{$\cal T$ solves $P$ with error $\eps$ \wrt $\mu$} if a correct answer is produced with probability at least $1-\eps$ when $(x,y)\sim\mu$.
The \e{$\eps$-error communication cost} of $P$ is the smallest possible message length of a protocol that solves $P$ with error $\eps$, and \e{$\eps$-error communication cost \wrt $\mu$} is defined similarly.
We say that the \e{bounded-error cost} of $P$ is at most $k$ if for any $\eps\in\RR$ its $\eps$-error cost is at most $k$.

Denote by $\RI[\eps](P)$ ($\RI[\mu,\eps](P)$) the classical one-way $\eps$-error communication cost of $P$ (\wrt $\mu$), and by $\RI(P)$ its bounded-error classical cost.
Denote by  $\QI[\eps](P)$, $\QI[\mu,\eps](P)$ and $\QI(P)$ the corresponding quantum analogs.

An important special case of relational communication problems are \e{functional} problems (partial or total).
The following theorem follows readily from \faketheoref{6} of~\cite{A04_Lim}:
\ntheo[t_Aa]{\cite{A04_Lim}}{For any functional two-party communication problem $F:X\times Y\to Z$, it holds that $\RI(F)\in\asO{\log(\sz Y)\QI[\eps](F)\log\QI[\eps](F)}$ for any constant $\eps<1/2$.}

\ssect[ss_single]{One-way communication when Bob receives no input}

In this section we present a new result in communication complexity that will be used later to prove \lemref{l_no_quasi}.

Consider a special case of one-way communication that we call \e{single-input mode}, where Bob receives no input.
Denote $\mathbf0\deq\set0$, and let $P\sbseq X\times\mathbf0\times Z$ be a communication task where Alice receives $x$ and sends a single message $m$ to Bob, who has to output $z\in Z$ based on the message $m$ alone.

This setting is not as trivial as it may appear at first glance.\fn
{It is important that we consider relational problems, for functions the single-input mode is indeed uninteresting.}
For instance, any communication problem with two-sided input $P\sbseq X\times Y\times Z$ has a single-input analogue $P'\sbseq X\times\mathbf0\times Z^Y$, where Bob has to produce a list of answers to the original $P$ \wrt all $y\in Y$.
Namely, let
\m{P_{\mu,\eps}'\deq\set[{\PR[y\sim\mu_x]{(x,y,z_y)\in P}\ge\eps}]{(x,0,(z_y)_{y\in Y})},}
where $\mu$ is a distribution on $X\times Y$ and $\mu_x$ is the marginal distribution of $B$ when $(A,B)\sim\mu$ and $A=x$.
Note that for any $\mu$ and $\eps\in\RR$, $\RI(P_{\mu,\eps}')\le\RI(P)$; on the other hand, by the Minimax theorem $\RI(P)=\Sup{\RI(P_{\mu,\eps}')}$, where the supremum is taken \wrt all possible $\mu$ and $\eps\in\RR$.

In other words, $P_{\mu,\eps}'$ is essentially as difficult to solve in the model of one-way \e{classical} communication as $P$ is.
Somewhat surprisingly, the same is not true in the case of quantum communication.
More generally, below we show that for any single-input communication task the quantum and the classical one-way costs are asymptotically the same.

\theo[t_QR]{For any relational two-party communication problem $P\sbseq X\times\mathbf0\times Z$, any distribution $\mu$ over $x\in X$ and $\eps\in\RR$, it holds that $\RI[\mu,\eps](P)\in\asO{\QI[\mu,\fr\eps3](P)}$.}

\crl[cr_QR]{For any $P\sbseq X\times\mathbf0\times Z$, it holds that $\RI(P)\in\asO{\QI(P)}$.}

\prf{By the Minimax theorem, for every $\eps$ there exists $\mu$ such that
$\RI[\eps](P)=\RI[\mu,\eps](P)$.}

If $P$ is a function then \crlref{cr_QR} is a very trivial special case of \theoref{t_Aa}.
On the other hand, \crlref{cr_QR} applies to the much more general case of relational problems, where a statement analogous to \theoref{t_Aa} provably does not hold.

\crl{There exists a functional two-party communication problem $F:W\to Z$ (where $W\sbs X\times Y$) and its single-input relational analogue $P\sbseq X\times\mathbf0\times Z^Y$, such that for some distribution $\mu$ over $X$ and $\eps\in\RR$ it holds that $\QI(F)$ is exponentially smaller than $\QI(P_{\mu,\eps})$.}

\prf{By \crlref{cr_QR}, this happens if and only if the gap between $\QI(F)$ and $\RI(F)$ is exponential.
An example of such $F$ was given in~\cite{GKKRW07_Expo}.}

\prf[\theoref{t_QR}]{Let $\Cl W$ be a valid \QI[\mu,\fr\eps3]-protocol of cost $m$ for $P$.

Let $A$ and $B$ be random variables taking the value of Alice's input $x\in X$ and Bob's answer $z\in Z$, respectively.
Assume $A\sim\mu$ and let $\mu^B$ be the corresponding distribution of $B$.
Conditional upon $A=x$, let $B\sim\mu_x^B$.
Denote by $\eps_x$ the probability that $\Cl W$ returns a wrong answer on input $x$, and let $Z_x\deq\set[(x,0,z)\in P]{z\in Z}$.
Then $\mu_x^B(Z_x)=1-\eps_x$.

We want to build an \RI[\mu,\eps]-protocol of cost \asO{m}.
Let $M\deq2^{\fr{16m}\eps}$, our classical protocol $\Cl W'$ will be as follows:
Alice and Bob use shared randomness to sample $M$ elements according to $\mu^B$, then Alice sends to Bob $\ceil{\log M}$ bits pointing to an element in the sampled set that belongs to $Z_x$, if one exists.
The communication cost of $\Cl W'$ is \asO{m}, and Bob's answer is correct if and only if an element from $Z_x$ has been sampled.

Denote
\m{\gamma_x\deq\KL{\mu_x^B}{\mu^B}=
\sum_z\mu_x^B(z)\log\fr{\mu_x^B(z)}{\mu^B(z)},}
and let us consider the probability that a randomly chosen $z\sim\mu^B$ belongs to $Z_x$.
From \lemref{l_substate},
\m{\PR[z\sim\mu_x^B]
{\fr{\mu_x^B(z)}{\mu^B(z)}\le2^{2\gamma_x+2}}\ge\fr12,}
and therefore
\m{\PR[z\sim\mu_x^B]
{\fr{\mu_x^B(z)}{\mu^B(z)}\le2^{2\gamma_x+2}
~\txt{and}~z\in Z_x}\ge\fr12-\eps_x.} 
Let $Z_x'\deq\set[\mu^B(z)\ge\mu_x^B(z)\tm2^{-2\gamma_x-2}]{z\in Z_x}$, then
\m[m_Z_x]{\mu^B(Z_x)\ge\mu^B(Z_x')\ge\mu_x^B(Z_x')\tm2^{-2\gamma_x-2}
\ge\l(\fr12-\eps_x\r)\tm2^{-2\gamma_x-2}.}

By Holevo's bound and the information processing principle,
\m{\E[x\sim\mu]{\gamma_x}=
\E[A=x]{\KL{\mu_x^B}{\mu^B}}=\I AB\le m,}
and from non-negativity of $\KL\dt\dt$ it follows that
\f{\PR[x]{\gamma_x>\fr[/]{7m}\eps}<\fr[/]\eps7.}
On the other hand, $\E[x]{\eps_x}\le\fr[/]\eps3$, and therefore
\f{\PR[x]{\eps_x>\fr[/]25}<\fr[/]{5\eps}6.}
From \bref{m_Z_x}, for sufficiently large $m$
\m{\PR[x]{\mu^B(Z_x)\le2^{-\fr{15m}\eps}}<\fr{41\eps}{42}.}
Therefore, $M$ random values sampled according to $\mu^B$ are likely to contain an element of $Z_x$, and the error probability of $\Cl W'$ is less than $\eps$ when $m$ is sufficiently large, as required.
}

\ssect{Connection to learnability of unspeakable concepts}

Let us see how \theoref{t_Aa} and \crlref{cr_QR} imply that our construction in \theoref{t_+} is tight.
First, let us see that no unspeakable \e{functional} concept class can be efficiently learned even in a quantum predictive learning model.
\lem[l_no_func]{Predictive learning of an unspeakable functional concept class is not possible from less than exponential amount of quantum information from the teacher, even by a computationally unlimited learner.}

\prf{Assume that for some functional concept class $\Cl F$ that is unspeakable, the following holds.
A teacher $\Cl T$ knows some $f_0\in\Cl F$, hidden from a learner $\Cl S$.
Then $\Cl T$ exchanges at most $k_q$ qubits with $\Cl S$.
Finally, $\Cl S$ is given some $x_0$ from the domain $X$ of the functions in $\Cl F$, and is able to compute $f_0(x_0)$ with confidence at least $5/6$.

Consider the following two-party communication task $\Cl G$.
Alice receives $f_0\in\Cl F$, Bob receives $x_0\in X$ and they have to output $f_0(x_0)$.
Clearly, $\QI[5/6](\Cl G)\le k_q$.

Let $k_c=\RI(\Cl G)$.
As $\Cl F$ is unspeakable, $k_c\in2^{\asOm n}$.
By \theoref{t_Aa}, $k_c\in\asO{n\tm k_q\log(k_q)}$, and so $k_q\in2^{\asOm n}$,
as required.}

Now we show that unspeakable concepts cannot be efficiently learned in the \e{quasi-predictive} (or standard) setting:
\lem[l_no_quasi]{Both standard and quasi-predictive learning of an unspeakable concept class is not possible from less than exponential amount of quantum information from the teacher, even by a computationally unlimited learner.}

\prf{It is enough to prove the statement only for quasi-predictive learning, and the standard model can be viewed as a special case.

Let $\Cl C$ be an unspeakable concept class consisting of relations over $X\times Y$, assume that it is learnable in the quasi-predictive model by a protocol of cost $k_q$.
Then there exists a protocol, according to which a teacher $\Cl T$ who knows some $\ell_0\in\Cl C$ exchanges at most $k_q$ qubits with a learner $\Cl S$ who doesn't know $\ell_0$.
Nevertheless, afterward $\Cl S$ is able to answer with sufficient confidence any number of testing questions regarding $\ell_0$.

For us it is enough to consider the testing phase where all possible $x\in X$ are asked (say, in the lexicographic order) and the learner responds with $(y_x)_{x\in X}$, such that
\m{\forall (\ell_0,x)\in \Cl C\times X:\PR{(x,y_x)\in\ell_0}\ge5/6,}
where the probability is taken \wrt possible runs of the learning protocol for the given $\ell_0\in\Cl C$.

Define a relational single-input communication problem $P_{\Cl C}\sbseq\Cl C\times\mathbf0\times Y^X$ as
\m{P_{\Cl C}\deq
\set[\sz{\set[(x,y_x)\in\ell_0]{x}}\ge\fr45\sz X]
{\llp\ell_0,0,(y_x)_{x\in X}\rrp}.}
The learning protocol for $\Cl C$ that we considered above can be turned into a \QI-protocol of cost $k_q$ for $P_{\Cl C}$ that is correct with probability $1-\aso1$ \wrt every $\ell_0\in\Cl C$, in particular $\QI(P_{\Cl C})\le k_q$.
By \crlref{cr_QR}, $\RI(P_{\Cl C})\in\asO{k_q}$.

Any \RI-protocol of cost $k_c$ for $P_{\Cl C}$ readily leads to an approximating class for $\Cl C$ of size $2^{k_c}$.
As $\Cl C$ is unspeakable, $k_c\in2^{\asOm n}$, where $n=\log\sz{X}$.
Therefore, $k_q\in2^{\asOm n}$, as required.
}

For simplicity, in the two proofs above we assumed distribution-free mode of learning, where the learner in the testing phase had to give correct answer to any $x\in X$ with high probability.
Distributional versions of \lemref[l_no_func]{l_no_quasi} can be proved similarly.

\sect{Open problems}

We demonstrated that efficient quantum predictive learning of an unspeakable relational concept class is possible.
The following questions seem interesting.

When we considered the limitations of quantum quasi-predictive learning (in the proof of \lemref{l_no_quasi}), we argued that a certain ``quasi-hypothesis'' of polynomial length can be extracted from an efficient quantum quasi-predictive learning algorithm.
But our construction does not rely upon the efficiency of the learning algorithm, and on the other hand, the quasi-hypothesis we construct cannot, in general, be efficiently evaluated.
It would be interesting to come up with a stronger argument that would ``preserve efficiency''; or otherwise, to give an example of an interesting quantum quasi-predictive learning algorithm.
Similar observations can be made \wrt our proof of \lemref{l_no_func}.
The transformation in~\cite{A04_Lim} is, in general, not efficient.
Are there interesting quantum predictive (or even quasi-predictive) learning algorithms for functional concepts?

In the above questions by ``interesting'' we meant quantum algorithms for learning a concept class that \e{admits} concise hypotheses, but only those that cannot be efficiently evaluated.
Observe that such ``quasi-unspeakable'' concept classes cannot be learned efficiently in any reasonable classical model (in the classical setting the equivalence between standard and predictive learning is efficiency-preserving).

Note that a trivial positive answer to these questions would follow, e.g., from an assumption that $\BQP\nsbse\P/poly$.
Therefore the goal should be to weaken the assumptions.

More generally, it would be interesting to see new examples of efficient quantum \mbox{(quasi-)} predictive learning of concept classes that are not efficiently learnable classically.
Such examples might be interesting even for models stronger than \pq\ (e.g., one may allow the learner to make \e{membership queries}).

\subsection*{Acknowledgments}
I am grateful to Rahul Jain for helpful discussions.
I have received many helpful comments from anonymous reviewers.

\bibliographystyle{alpha}

\MyComment{Spell-check}

\bib

\end{document}